\documentclass[aps, prl, showpacs, twocolumn, 10pt, superscriptaddress, a4paper, floatfix]{revtex4-1}

\usepackage{graphicx}        % standard LaTeX graphics tool
\usepackage{epstopdf}
\usepackage{CJK}
\usepackage{pifont}
\usepackage{bm}
\usepackage{amssymb}
\usepackage{amsmath}
\usepackage{siunitx}
\usepackage[usenames,dvipsnames]{xcolor}
\usepackage[linkcolor=violet,citecolor=blue,colorlinks=true,urlcolor=purple]{hyperref}

\newcommand{\SM}{{Supplemental Material \cite{SMHanus2019}}}

\newcommand{\ssg}{$1s\sigma_g$}
\newcommand{\psu}{$2p\sigma_u$}
\renewcommand{\vec}{\textbf}

\begin{document}
%\tableofcontents
\begin{CJK*}{UTF8}{gbsn}

\title{Sub-femtosecond tracing of molecular dynamics during strong-field interaction}
%\title{Draft: The Molecular Attoclock: Sub-cycle Control of Electronic Dynamics during H2 Double Ionization}
%\title{Control over fragmentation reactions of polyatomic molecules with impulsive alignment}

\author{V\'{a}clav\,Hanus}
\email[]{vaclav.hanus@tuwien.ac.at}
\affiliation{Photonics Institute, Technische Universit\"at Wien, 1040 Vienna, Austria, EU}

\author{Sarayoo\,Kangaparambil}
\affiliation{Photonics Institute, Technische Universit\"at Wien, 1040 Vienna, Austria, EU}

\author{Seyedreza\,Larimian}
\affiliation{Photonics Institute, Technische Universit\"at Wien, 1040 Vienna, Austria, EU}

\author{Martin Dorner-Kirchner}
\affiliation{Photonics Institute, Technische Universit\"at Wien, 1040 Vienna, Austria, EU}

\author{Xinhua\,Xie (谢新华)}
%\author{Xinhua\,Xie}
\affiliation{Photonics Institute, Technische Universit\"at Wien, 1040 Vienna, Austria, EU}
\affiliation{SwissFEL, Paul Scherrer Institute, 5232 Villigen PSI, Switzerland}

\author{Markus\,S.\,Sch\"{o}ffler}
\affiliation{Institut f\"ur Kernphysik, Goethe-Universit\"at Frankfurt, 60438 Frankfurt, Germany, EU}

\author{Gerhard\,G.\,Paulus}
\affiliation{Institute for Optics and Quantum Electronics, Friedrich-Schiller-Universit\"at Jena, 07743 Jena, Germany, EU}

\author{Andrius\,Baltu\v{s}ka}
\affiliation{Photonics Institute, Technische Universit\"at Wien, 1040 Vienna, Austria, EU}

\author{Andr\'{e} Staudte}
\email[]{andre.staudte@nrc-cnrc.gc.ca}
\affiliation{Joint Laboratory for Attosecond Science of the National Research Council and the University of Ottawa, Ottawa, Ontario K1A 0R6, Canada}

\author{Markus\,Kitzler-Zeiler}
\email[]{markus.kitzler@tuwien.ac.at}
\affiliation{Photonics Institute, Technische Universit\"at Wien, 1040 Vienna, Austria, EU}

\begin{abstract}
%The Born-Oppenheimer approximation codifies the assumption that the nuclear motion in molecules is slow on the time scale of electronic motion. When an electron is suddenly removed from a molecule through multiphoton ionization by an intense, infrared laser pulse, the molecule finds itself in a superposition of ionic states. The electronic eigenstates of the ion can often be treated as an incoherent sum, whereas the nuclear eigenstates usually form a wavepacket, which is launched on the changed potential energy landscape.
We introduce and experimentally demonstrate a method, where the two intrinsic time scales of a molecule, the slow nuclear motion and the fast electronic motion, are simultaneously measured in a photo-electron photo-ion coincidence experiment. In our experiment, elliptically polarized, 750~nm, 4.5~fs laser pulses were focused to an intensity of $9\times10^{14}\mathrm{W/cm}^2$ onto H$_2$. Using coincidence imaging, we directly observe the nuclear wavepacket evolving on the \ssg{} state of H$_2^+$ during its first roundtrip with attosecond temporal and picometer spatial resolution. The demonstrated method should enable insight into the first few femtoseconds of the vibronic dynamics of ionization-induced unimolecular reactions of larger molecules.
%
%An important aspect of strong-field laser-molecule interaction is that the electronic and nuclear dynamics in molecules depend on each other. The laser electric field rips off electrons that participate on the molecular bond leaving behind an ionized molecule with its molecular bond in motion. The motion rearranges the molecule, changes the electronic structure and the dynamics of next ionization steps accordingly.
%Here we demonstrate a technique that allows to observe the fast sub-cycle electronic dynamics in relation to the slower nuclear dynamics and in turn allows to trace the motion of nuclei of the molecule during its early stage which is not accessible by pump-probe techniques. The technique is applied to tracing the vibrational wavepacket upon single ionization of hydrogen. The measurements reveal that it takes place on the  H$_2^+$ \ssg{} energy level.
\end{abstract}

%\pacs{33.80.Rv, 42.50.Hz, 82.50.Nd}

% 33.80.Rv 	Multiphoton ionization and excitation to highly excited states (e.g., Rydberg states)

% 42.50.Hz 	Strong-field excitation of optical transitions in quantum systems; multiphoton processes; dynamic Stark shift (for multiphoton ionization and excitation of atoms and molecules, see 32.80.Rm, and 33.80.Rv, respectively)

% 82.30.Qt 	Isomerization and rearrangement
% 82.20.Kh 	Potential energy surfaces for chemical reactions
% 82.37.Np 	Single molecule reaction kinetics, dissociation, etc.
% 82.50.Nd 	Control of photochemical reactions
% 82.50.Pt 	Multiphoton processes

\maketitle
\end{CJK*}

%%%%%%%%%%%%%%%%%%%%%%%%%%%%%%%%%%%%%%%%%%%%%%%%%%%%%%%%%%%%%%%%%%%%%%%%%%%%%%%

%----------------------------------------------
%------------------------- general introduction
%----------------------------------------------

%Timing of electron emission from molecules can have a significant influence on the molecular dynamics and can affect the outcome of the laser-molecule interaction. It has been shown for example that changing the timing of the electron removal by varying the pulse duration can change the number of fragments of ethylene \citep{Xie2015c}. In experiments with acetylene \citep{Erattupuzha2017c} the stretch motion of CH bonds triggered by the initial ionization steps influences the final ionization degree.

%\cmmt{Figure captions not yet correct.}

Molecular fragmentation and isomerization processes are of fundamental importance in nature. These nuclear rearrangement processes are initiated and ultimately determined by electronic dynamics that can be influenced by precisely timed distortions of the electronic structure with the electric field of strong, ultrashort laser pulses \cite{Kling2013, Alnaser2017, Xie2012_CE, Xie2014_PRX}.
For example, fine variations of the delay between successive ionization events can determine the number of moieties produced during fragmentation of polyatomic molecules \cite{Xie2015}. %, Erattupuzha2016}.
To reveal the dynamics underlying these processes, it is necessary to apply probing techniques that are sensitive to both, the fast electronic dynamics that may take place on attosecond time scales and the slower nuclear motion taking place on the femtosecond time scale.
%While it is possible to probe even the fastest vibrational dynamics of H$_2^+$
While even the fastest vibrational wavepacket in H$_2^+$ has been probed successfully shortly \textit{after} its creation using a pump-probe scheme with near-single-cycle pulses \cite{Ergler2006_D2_nuclear_wavepacket, Xu2015c}, it is, however, a major obstacle to trace the molecular dynamics \textit{during} the first few femtoseconds of the interaction of a molecule with a strong laser pulse.
%The reason is the limit in temporal resolution of pump-probe methods to about the pulse duration, which for laser pulses in the visible and near/mid-infrared cannot be sub-femtoseconds. Attosecond pulses in the extreme ultraviolet, however, probe unfortunately different dynamics \cite{Sansone2010, Trabattoni2015, Nabekawa2015}.

In this Letter, we introduce and experimentally demonstrate a method that is capable of tracing simultaneously electronic and vibrational wavepacket dynamics in a fragmenting molecule on sub-femtosecond times.
The method combines the sub-cycle sensitivity inherent to angular streaking that has been applied to studying ionization dynamics in atoms \citep{Maharjan2005, Eckle2008a, Pfeiffer2011_Nature_Phys, Schoffler2016, Wustelt2017} and molecules \cite{Wu2012b, Wu2012, Wu2013, Gong2014}, with the high structural sensitivity of Coulomb explosion imaging \cite{Stapelfeldt1998, Niikura2003,Legare2003,Legare2005,Erattupuzha2017} to resolve vibrational motion.
Our work is inspired by earlier attempts \cite{Staudte2005} of constructing such a method that has been called \textit{The Molecular Clock}.

\begin{figure*}[htb]
  \centering
  \includegraphics[width=0.95\textwidth]{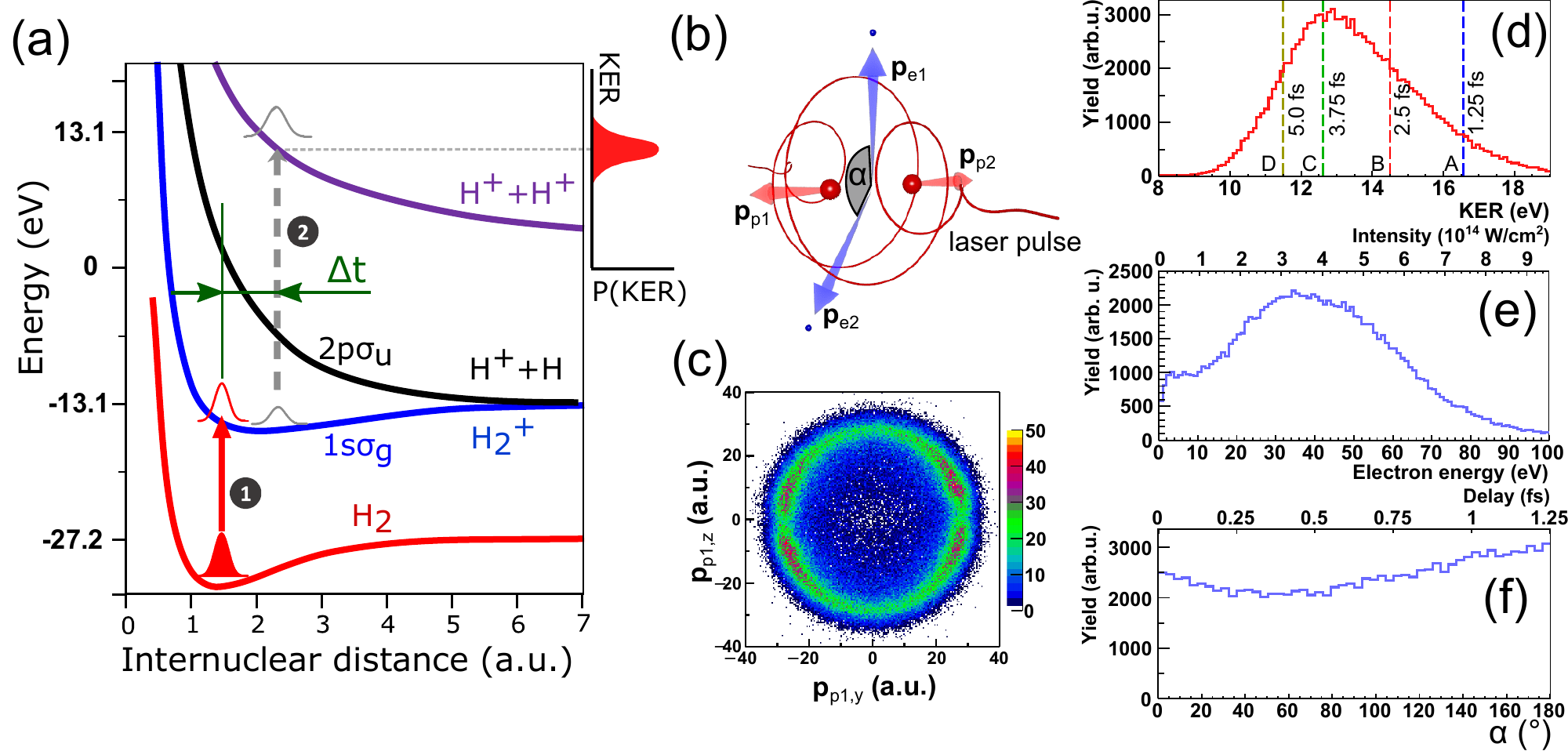}
  \caption{(a) Schematics of the relevant potential energy surfaces of H$_2$. Indicated by arrows are the two ionization steps (\ding{182} and \ding{183}) that are delayed with respect to each other by $\Delta t$.
  (b) Cartoon of the electron and ion momentum vectors after double ionization of H$_2$ in an intense elliptically polarized few-cycle laser pulse.
  (c) Distribution of the proton momentum from intra-pulse double ionization in the polarization plane. The histogram is filled with two detected protons which satisfy a center-of-mass momentum $<$10~a.u. The minimum around $p_{p1,z} = 0$~a.u. is due to the dead time of the detector.
  (d) Kinetic energy release (KER). The lines A-D label certain values of $\Delta t$.
  (e) Photoelectron energy distribution. Upper axis: corresponding laser intensity.
  (f) Relative emission angle between the two photoelectrons. Upper axis:  corresponding ionization delay on a sub-cycle scale.}
  \label{fig1}
\end{figure*}

We investigate the applicability of this concept
by tracing the nuclear and electronic dynamics in $\text{H}_2^+$ triggered by the emission of an electron over several femtoseconds.
Upon ionization of $\text{H}_2^+$ various scenarios can take place \cite{Ibrahim2018}. The simplest
one is that a vibrational wavepacket is created on the \ssg{} energy level following the Franck-Condon (FC) principle \cite{Niikura2002}, see Fig.~\ref{fig1}(a). Even for this simplest of all cases it has been shown that the very fast nuclear motion taking place during the ionization event can lead to deviations from the population distribution of vibrational states predicted by the FC principle \cite{Urbain2004}. In another scenario, shake-up excitations can lead to the population of vibrational states not only on the \ssg{} but also on the \psu{} level \cite{Litvinyuk2005}. In recent photoelectron holography measurements evidence for subcycle population transfer to \psu{} was found \cite{Haertelt2016}.
As the vibrational dynamics proceeds after ionization in the presence of the strong laser field, three- or five-photon resonant excitations, can lead to partial population transfer between the \ssg{} and \psu{} levels \cite{Kling2013, Alnaser2017}.
We show that we can achieve temporal and spatial resolutions of a few tens of attoseconds and about 1\,pm -- sufficient to disentangle these different scenarios.

To obtain attosecond temporal resolution we employ angular streaking \citep{Maharjan2005, Eckle2008a} illustrated in Fig.~\ref{fig1}(b). Angular streaking exploits the direct mapping of electron momentum to the instantaneous laser electric field at the time of ionization. In elliptically polarized light the electric field vector $\vec{E}(t)$ completes a full rotation within one period of the field, $T=2\pi/\omega$, where $\omega$ is the frequency of the light. This rotation serves as the minute hand of a clock: The ionization phase within a laser cycle $\varphi=\omega t_i$ is mapped into the emission angle of the photoelectron via the relation $\vec{p}_e = -\vec{A}(t_i)$, valid within the strong-field approximation \cite{Faisal1973, Reiss1980}, where the laser vector potential $\vec{A}(t)$ is connected to the laser electric field by $\vec{A}(t)=-\int_{-\infty}^{t}\vec{E}(t')dt'$. Thus, measurement of the electron emission angle in the laboratory frame determines the ionization time $t_i$ within one cycle. In addition, the magnitude of the emitted electron's momentum vector $|\vec{p}_e|$ is proportional to the instantaneous field strength and thereby provides a measure about the ionization time within the pulse envelope \cite{Pfeiffer2011_Nature_Phys}.

The hand of the clock on the vibronic time scale is the kinetic energy release (KER) of the protons following double ionization. It has been shown that the kinetic energy released during Coulomb explosion of a molecule into two ionic fragments is a sensitive measure of the distance $R$ where the repulsive Coulombic potential is populated \cite{Vager1989, Stapelfeldt1998, Chelkowski1999, Legare2005, Katsuki2006}.
In H$_2$ the kinetic energy released during Coulomb explosion is $\text{KER} =
\frac{1}{2 m_p}\left( \vec{p}_{p1}^2 + \vec{p}_{p2}^2 \right)$, where $\vec{p}_{p1}$ and $\vec{p}_{p2}$ are the proton momenta and $m_p$ is the proton mass.
The mapping of internuclear distance $R = \frac{1}{\text{KER}}$ to kinetic energy release is very precise for H$_2$ \cite{Niikura2003, Weber2004}. The first ionization event at time $t_1$ initiates a H$_2^+$ nuclear wavepacket on the \ssg{} ground state. The  wavepacket propagates on the light-induced potential energy surfaces and is projected onto the Coulombic potential energy curve by the second ionization event at time $t_2=t_1+\Delta t$, see Fig.~\ref{fig1}(a). Measuring the momenta of the two protons provides us with the internuclear distance $R$ at which the second ionization step happened. Hence, if the two ionization events are confined to within the first roundtrip of the nuclear wavepacket on the \ssg{} potential we can establish a correlation between KER and $\Delta t$.

For the experiment we used reaction microscopy \cite{Doerner2000}. We measured the three-dimensional momentum vectors of two protons in coincidence with two electrons emerging from the interaction of H$_2$ molecules with elliptically polarized pulses with a broad spectrum centered around 750\,nm (oscillation period $T=2.5$\,fs), a full-width at half maximum (FWHM) duration of 4.5~fs in intensity
and a peak intensity of \SI{9e14}{\watt\per\square\cm}.
The experimental apparatus consists of a two-stage arrangement to provide an internally cold ultrasonic gas jet of hydrogen, and an interaction chamber with an ultra-high vacuum ($1.3\times10^{-10}$mbar).
%The laser beam, focused to a peak intensity of \cmmt{insert intensity}, intersected with an ultrasonic jet of H$_2$ in the interaction chamber (background pressure $1\times10^{-10}$\,mbar). % using a spherical silver mirror with a focal length of 60\,mm.
%When a hydrogen molecule interacts with a laser pulse, it can be singly ionized and forms \h2p which may dissociate into a proton and a hydrogen atom.
Electrons and ions were guided by weak magnetic (12\,G) and electric (21\,V/cm) fields along the spectrometer axis ($z$-direction) to two position and time sensitive multi-hit detectors.
%The few-cycle laser pulses were generated by spectral broadening of 25-fs pulses from a Titanium-Sapphire laser amplifier system that were sent into a hollow-core fiber filled with neon gas and compressed using pairs of chirped mirrors.
%The duration of the pulses was 4.5\,fs [full width at half maximum (FWHM) in intensity].
% were obtained and the pulse duration was monitored during the experiment using a phase-meter device \cite{Sayler2011a, Schoffler2016, Xie2015}.
Further details on the reaction microscope can be found in Refs.\,\cite{Xie2012_CE, Zhang2014a, Xie2017b} and on the optical setup in Ref.\,\cite{Schoffler2016} as well as in \SM{}.

\begin{figure*}[htb]
  \centering
  \includegraphics[width=0.95\textwidth]{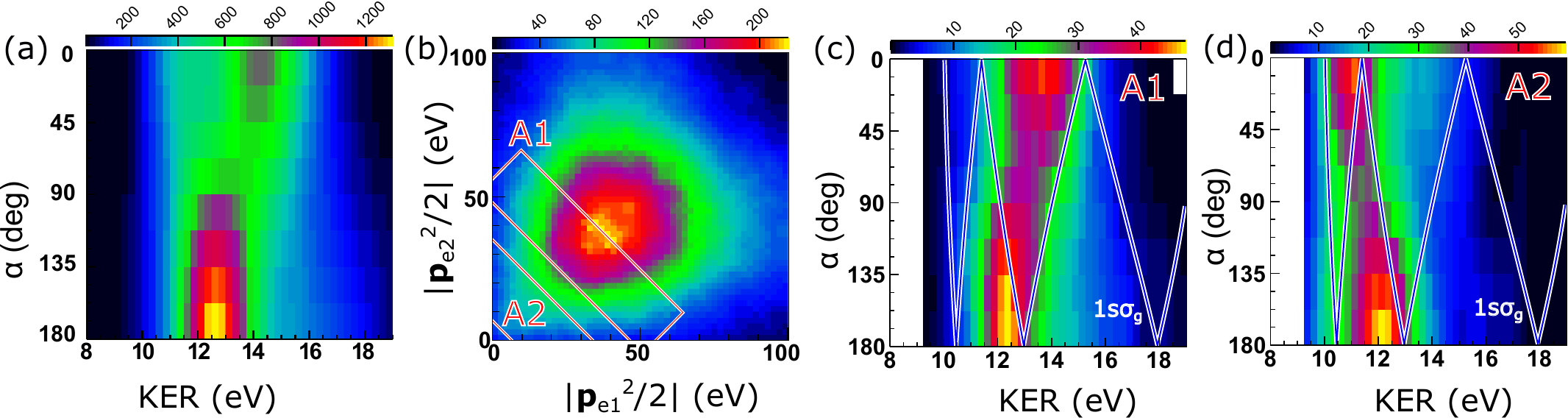}
  \caption{(a) Double ionization yield as a function of kinetic energy release (KER) and the relative electron emission angle, integrated over the photoelectron energy.
  (b) Energy distribution of the two emitted photoelectrons (symmetrized about the diagonal). The marked areas \textit{A1, A2} correspond to regions of constant sum energy. Most of the photoelectrons are created with similar energies which indicates that the two-electron emission is symmetrically distributed around the laser pule peak.
  (c) Measured fragmentation yield as a function of $\alpha$ and KER obtained when the momenta of the two emitted electrons lie within range \textit{A1} indicated in (b).
  (d) Same as (c) but for electron momenta within range \textit{A2}.
The thin line in (c) and (d) shows the classical expectation value obtained for vibrational motion on the \ssg{} potential energy curve (see text and \SM{} for details). }
  \label{fig2}
\end{figure*}

Fig.~\ref{fig1}(c) shows the proton momentum distribution in the laser  polarization plane for the double ionization pathway. The anisotropy of the proton distribution indicates the alignment of the polarization ellipse in the laboratory frame. In elliptical light, ionization preferentially takes place twice during the optical cycle \cite{Eckle2008}. Hence, sequential double ionization in elliptical light occurs with a delay between the two ionization steps of $\Delta t = nT/2; n=0,1,2,\hdots$.
To calibrate the zero of our time-scale we turn to the KER, shown in Fig.~\ref{fig1}(d). For instantaneous double ionization, i.e., $\Delta t=0$, the H$_2$ groundstate wavefunction would be projected vertically onto the Coulomb repulsion curve, to yield a KER distribution centered at 19~eV with a FWHM of 2~eV \cite{Weber2004}. Thus, based on Fig.~\ref{fig1}(d) we can exclude instantaneous double ionization in our experiment.

The observed KER distribution from 10~eV to 18~eV can be related to a timescale, assuming nuclear wavepacket propagation on a given potential energy surface. To this end, we computed the nuclear-stretch motion in between the ionization times $t_1$ and $t_2=t_1+\Delta t$ by solving Newton's equations on the \ssg{} energy curve \cite{Sharp1970}; see \SM{} for details on the simulations. The simulation shows that our experiment covers a range of ionization delays starting from about $\Delta t =0.5T$ ($\textrm{KER}\approx 18$\,eV) to roughly $\Delta t=3T$ ($\textrm{KER}\approx10$\,eV), with a maximum of the double ionization probability around $\Delta t=1.5T$ ($\textrm{KER}\approx 13$\,eV).
Smaller $\Delta t$-values correspond to smaller internuclear distances, $R$, where the ionization potential is greatly increased and double ionization probability is accordingly suppressed, see Fig.\,\ref{fig1}(a).
Double ionization delays of $\Delta t \lesssim 0.5T$ require both the increase of the pulse peak intensity and a pulse duration even shorter than the 4.5-fs pulses used here (\SM{}).

The energy of each photoelectron, shown in Fig.~\ref{fig1}(e), labels the light intensity at the instant of ionization. The peak at 40~eV corresponds to an ionization intensity of about $3.8 \times 10^{14} \mathrm{W/cm}^{2}$. Therefore, the majority of double ionization events does not occur at the peak of the pulse, in agreement with the prediction of the double ionization delay based on the KER.
Sub-cycle sensitivity, finally, is obtained from the relative angle $\alpha =\text{ang}\left(\vec{p}_{e1},\vec{p}_{e2}\right)$ between the two photoelectron momenta $\vec{p}_{e1}$ and $\vec{p}_{e2}$, shown in Fig.~\ref{fig1}(f). By virtue of the relation $\vec{p}_{e1,e2} = -\vec{A}(t_{1,2})$ the relative angle $\alpha \in [0^{\circ},180^{\circ}]$ directly measures $\Delta t$ within one laser half-cycle. The minimum at $\alpha \approx 90^{\circ}$ can be attributed to the ellipticity of the pulse. The relative emission angle peaks at $\alpha=180^{\circ}$, thus indicating a preferential double ionization at a delay of half-integer optical cycles, consistent with the $\Delta t=1.5T$ deduced from the maximum of the KER distribution.

However, the estimates based on the isolated observables in Fig.~\ref{fig1}(d-f) do not yield a true sub-cycle resolution of the coupled electronic and nuclear dynamics. To obtain this, we will in the following examine the relation between KER, photoelectron emission angle and energy to demonstrate how the sub-cycle dynamics of the nuclear wavepacket can be accessed by correlating these three observables.
In Fig.~\ref{fig2}(a) we show the correlation between the two main hands of the molecular clock: the relative emission angle $\alpha$ of the photoelectrons and the KER of the protons. A previously indistinguishable maximum in the double ion yield at a KER of 14.5~eV is associated with a relative emission angle of zero degrees, i.e., the parallel emission of the two photoelectrons. The main maximum at a KER of 12.6~eV is slightly lower than in Fig.~\ref{fig1}(d) and associated with anti-parallel electron emission. Finally, a third peak at a KER of 11.5~eV and parallel emission is weakly distinguishable.

Thus, by correlating $\alpha$ with KER, we obtain a powerful parametric framework for representing the coupled electron and nuclear dynamics taking place in between the two ionization steps.
For example, if the \psu{} energy level is populated by shake-up during the first ionization step or also at later times by resonant transitions, the corresponding parametric curves in the $\alpha$-KER frame of reference will look distinctively different from the one corresponding to vibrational motion on the \ssg{} level; see  \SM{} for a comparison of curves obtained for different scenarios.
Therefore, the ionization-fragmentation dynamics can be studied by comparison of the measured fragmentation yield to a simulated curve for a given scenario, e.g., for motion only on the \ssg{} energy level.

Further insight into the molecular dynamics within the duration of the pulse can be obtained when the magnitudes of the photoelectron momenta $|\vec{p}_{e1,e2}|$ are analyzed. $|\vec{p}_{e1,e2}|$ provide an additional time reference linked to the fast rise time of a few-cycle pulse's envelope \cite{Pfeiffer2011_Nature_Phys}. Gating on $|\vec{p}_{e1,e2}|$, thus allows selecting a range of ionization delays $\Delta t$. As a result, it becomes possible to obtain sub-cycle traces of the molecular dynamics for certain ranges of delays $\Delta t$ in the two-electron emission and, moreover, as we will show further below, even attosecond snapshots of the propagating vibrational wavepacket.

For the purpose of the following analysis we show the energies of both photoelectrons in Fig.~\ref{fig2}(b). The highest double ionization probability is found for similar momenta of both electrons. Hence, the highest double ionization yield is from events where both electrons are emitted symmetrically around the peak of the pulse envelope. For example, the region \textit{A1} indicated in the correlated energy spectrum Fig.~\ref{fig2}(b)  corresponds to situations where both electrons are emitted within $\Delta t= 1.2T-1.7T$. Region \textit{A2} at smaller energies corresponds to emissions with longer delay, $\Delta t= 2.3T-3.2T$.

The capabilities opened up by the selection of the emission time-window based on $|\vec{p}_{e1,e2}^2/2|$ are demonstrated in Figs.~\ref{fig2}(c,d). In these figures we show the measured distribution of the fragmentation yield in the $\alpha$-KER plane for the two regions of electron energies \textit{A1} and \textit{A2} in Fig.~\ref{fig2}(b).
Accordingly, the corresponding ranges of the yield distribution in the $\alpha$-KER plane in Figs.~\ref{fig2}(c) and (d) show the vibrational wavepacket evolution approximately in the time ranges of $1.2T-1.7T$ and $2.3T-3.2T$ after the first ionization step, respectively.
A movie obtained for arbitrary selections of $|\vec{p}_{e1,e2}|$ is available as \SM{}.
The measured distributions in Figs.~\ref{fig2}(c,d) are compared to the simulated curve obtained for wavepacket motion on the \ssg{} energy level (described in \SM{}).
The measured distributions show no signs of shake-up excitation that would appear for KER$>19$~eV. Likewise, they show negligible yield in the range KER$<10$~eV that corresponds to resonant excitation to the \psu{} energy curve. Overall, the measured distributions agree well with the simulated curve for nuclear motion on the \ssg{} curve, although there are some small deviations visible for the smaller emission delays $\Delta t$, cf. Fig.~\ref{fig2}(c).

As $\alpha=0-180^\circ$ corresponds to one laser half cycle, by limiting $\alpha$ to small intervals, attosecond snapshots of the KER distribution can be extracted from these traces.
By virtue of the Coulomb law, KER$=1/R$ can be converted to internuclear distance $R$.
Thus, it becomes possible to obtain snapshots of the absolute value of the vibrational wavepacket, $|\chi(R,\Delta t)|^2$, for very short time-intervals  around certain values of $\Delta t$.

\begin{figure}[tb]
  \centering
  \includegraphics[width=\columnwidth]{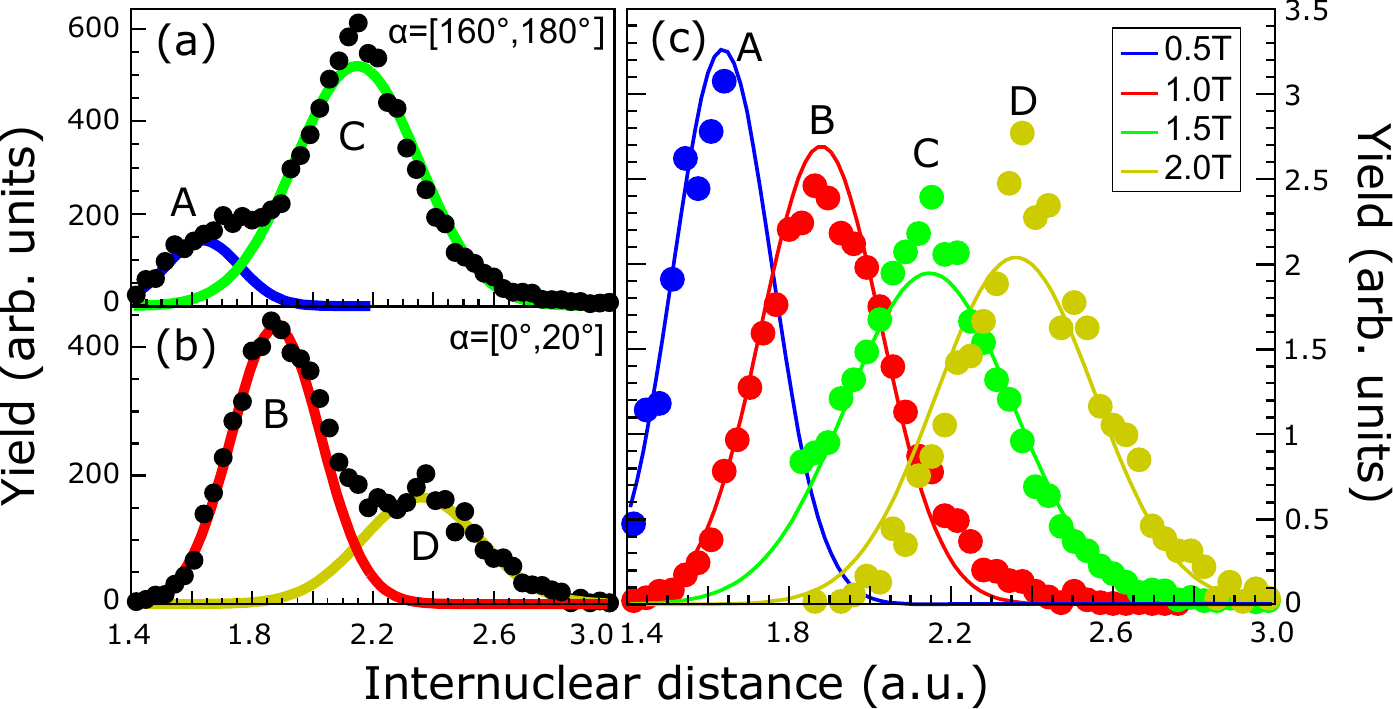}
  \caption{Measured fragmentation yield (dots) over $R$ for ionization events separated by an even (a) and odd (b) number of half cycles. (a) The distributions result from the selection  $\alpha=[160^\circ,180^\circ]$, corresponding to ionization events taking place within a range of $\pm 70$\,as around $\Delta t = 2n\frac{T}{2}, n=0,1,2$. (b) Same as (a) but for $\alpha=[0^\circ,20^\circ]$, corresponding to $\pm 70$\,as around $\Delta t = (2n+1)\frac{T}{2}, n=0,1,2$. The colored lines labeled A-D are Gaussian fits to the data. The labels A-D correspond with those in Fig.~\ref{fig1}.
  (c) Measured fragmentation yield (dots) from (a,b) but with additional restrictions on the magnitude of the electron momentum (see \SM{} for details) in comparison with the Gaussian fits from (a,b). The areas of the distributions A-D were normalized to one.}
  \label{fig3}
\end{figure}

To obtain $|\chi|^2$ we select slices in the $\alpha$-KER distribution for $\alpha = [0^\circ,20^\circ]$ and $\alpha = [160^\circ,180^\circ]$ corresponding to intervals of $\Delta t$ of about 140\,as. The yield distributions obtained by these selections are plotted in Fig.\,\ref{fig3}(a,b) as a function of $R$ together with Gaussian fits. The Gaussian fits agree well with yield distributions obtained by more sophisticated selections (detailed in \SM{}) that also involve gating on $|\vec{p}_{e1,e2}|$ as described above, see Fig.\,\ref{fig3}(c).
The four distributions in Fig.\,\ref{fig3}(c) constitute snapshots of $|\chi(R,\Delta t)|^2$ with an uncertainty of about $\pm 70$\,as around $\Delta t = 0.5T,1T,1.5T,2T$ and with a spatial resolution of about 0.02\,a.u. (about 1\,pm).

The equilibrium internuclear distance of H$_2$ is 1.4\,a.u.
Fig.\,\ref{fig3}(c) shows that after half a laser cycle ($\Delta t=1.25$\,fs) the H$_2^+$ nuclear wavepacket has propagated in distance by about 0.2\,a.u.
For larger values of $\Delta t$, our measurement reveals how the wavepacket progressively moves to larger internuclear distances and spreads in space. Compared with previous experiments that probed the nuclear motion of H$_2^+$ after strong-field laser ionization \cite{Niikura2003}, our measurement is, to the best of our knowledge, the first one to reveal not only the position but also the shape of the wavepacket. Additionally, our measurement probes the wavepacket on a finer grid of $\Delta t$ and with significantly higher temporal resolution (i.e., uncertainty in $\Delta t$).

%----------------------------------------------
%------------------------- summary and discussion
%----------------------------------------------

In conclusion, we introduced a new method, that allows tracing molecular dynamics on sub-femtosecond times \textit{during} strong-field interaction following field-ionization. The method exploits the rotation of the electric field vector of elliptically polarized light as an attosecond temporal reference, and the dependence of the ion fragment energy on the molecular geometry as a clock for nuclear motion.
Although demonstrated here for H$_2$, for which the nuclear clock shows a $1/R$ dependence, this is not a pre-requisite of the method; any monotonic dependence of fragment energy on nuclear geometry is suitable. We therefore expect a wide applicability of the presented method and envision that it can also be applied to reasonably fast dissociative few-particle fragmentation channels in polyatomic molecules, as many of them fulfill this requirement.
We highlight the high temporal resolution and spatial sensitivity of the method by demonstrating tracing of the vibrational wavepacket evolution in H$_2^+$ with a temporal uncertainty of about $\pm70$\,as and a spatial resolution of about 1\,pm.
The resolution of this technique can be expected to be further improved by exploitation of the carrier-envelope phase of few-cycle laser pulses \cite{Schoffler2016}.

%------------------------- Acknowledgments

\acknowledgments
This work was financed by the Austrian Science Fund (FWF), Grants No.
P28475-N27,  % BOND Markus
%P25615-N27,  % Atto EWP Interferometry Xinhua
and P30465-N27.  % O2 production Xinhua

%%%%%%%%%%%%%%%%%%%%%%%%%%%%%%%%%%%%%%%%%%%%%%%%%%%%%%%%%%%%%%

%\bibliography{refs_markus,refs_vaclav}
%\bibliography{refs_markus,Supp_Mat}


\end{document}

{\bf \noindent \sffamily \huge \color{titlecolor} Sub-femtosecond tracing of molecular dynamics during strong-field interaction}

\vspace{0.8cm}

{\bf \noindent \sffamily \Large Supplemental Material}

\vspace{0.8cm}

{\bf \noindent \large V\'{a}clav\,Hanus, Sarayoo\,Kangaparambil, Seyedreza\,Larimian, Martin\,Dorner-Kirchner, Xinhua\,Xie, Markus\,S.\,Sch\"{o}ffler, Gerhard\,G.\,Paulus, Andrius\,Baltu\v{s}ka, Andr\'{e} Staudte, Markus\,Kitzler-Zeiler}

%%%%%%%%%%%%%%%%%%%%%%%%%%%%%%%%%%%%%%%%%%%%%%%%%%%%%%%%%%%%%%%%%%%%%%%%%%%%%%%

\section{Experimental details\label{sec:exp_details}}

%\textbf{Laser intensity and ellipticity calibration} 
%

Peak intensity and ellipticity were calibrated by analyzing the electron momentum distribution obtained for single ionization of helium with the same pulse as in the experiment on H$_2$, using the method described in Ref.~\citen{Smeenk2011}. 
By this approach we obtained the laser intensity in the interaction region as \SI{9e14}{\watt\per\square\cm}. The ellipticity, defined as the ratio of the electric field strengths perpendicular and parallel to the main axis of the polarization ellipse, $E_\perp/E_\parallel$ was \num{0.85}. 
The direction of the polarization ellipse was determined from the angular distribution of protons emitted by Coulomb explosion of doubly charged hydrogen, H$_2^{2+}$. The rotation of the molecule during the ionization and Coulomb explosion processes was neglected given the short pulse duration used in the experiment.
%The proton momentum distribution is assumed to be peaked along the direction of the major axis of the polarization ellipse. 
%
To minimize smearing of the laser intensity along the propagation direction of the laser beam (coordinate $x$), the gas jet (propagating along $y$) was cut by adjustable razor blades to about 10\,$\mu$m, much shorter than the Rayleigh length of the laser beam ($\approx$200\,$\mu$m). %By this and by placing the focus of the laser beam slightly before the narrow jet, not only intensity smearing was minimized, but also averaging of CEP due to the Gouy phase shift was avoided. 
The duration of the pulses ($\approx$\SI{4.5}{\fs}) and their CEP-values (not analyzed in the present work) were obtained with a phase-meter device \cite{Sayler2011a}.

\begin{figure}[ht]
  \centering
  \includegraphics[width=0.5\columnwidth]{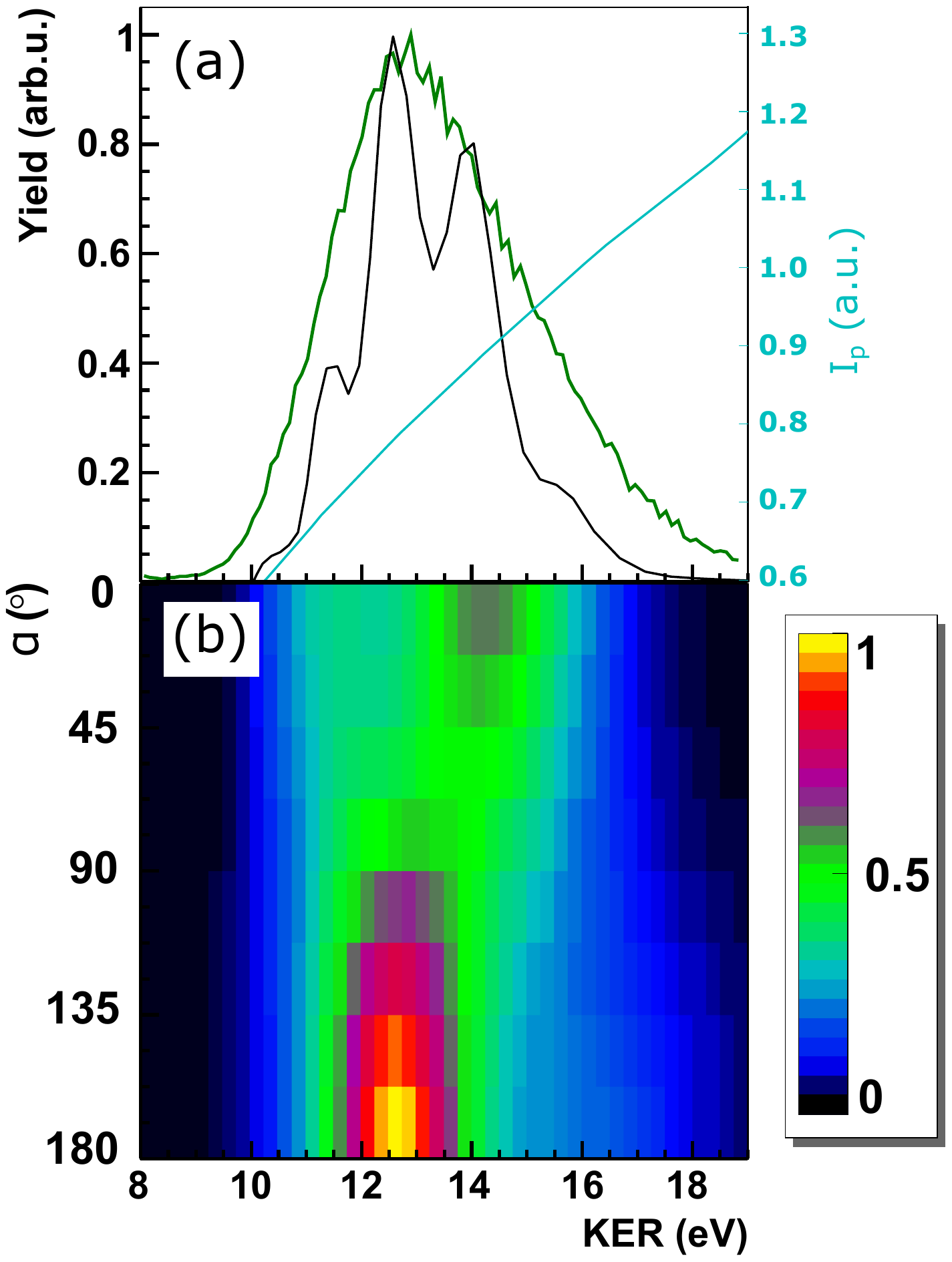}
  \caption{(a) Measured distribution of the KER (green) in comparison with simulated data (black).  The light blue line (ordinate on the right applies) depicts the dependence of the ionization potential on KER used in the simulation. See Sec.~\ref{sec:simulations_ionization_dynamics_KER} for details. (b) The measured KER distribution from (a) resolved in the relative emission angle of the two electrons, $\alpha$.}
  \label{fig1}
\end{figure}

In our experiments we use a reaction microscope to measure the momentum vectors of the two electrons, $\vec{p}_{e1,e2}$ and two protons $\vec{p}_{\text{p1},\text{p2}}$ ejected during fragmentation of doubly charged hydrogen, H$_2^{2+}$. In some cases the dead time of the detectors inhibits the detection of both electrons and only one electron is detected. In these cases we calculated the momentum of the missed  electron by exploiting momentum conservation between the momentum of the other electron and the ions, $\vec{p}_{e2} = -(\vec{p}_\text{p1} + \vec{p}_\text{p2} + \vec{p}_{e1})$. 
The kinetic energy released during the Coulomb explosion of the two protons after double ionization is given by $\text{KER}=\frac{1}{2m_p}\left( \vec{p}_\text{p1}^2 + \vec{p}_\text{p2}^2 \right)$, with $m_p$ being the proton mass ($\approx 1836$ in atomic units). The measured distribution of KER is shown in Fig.~\ref{fig1}(a).
From the electron momentum vectors we calculate the emission angle $\alpha$ between the two electrons using the scalar product, $\cos(\alpha)=\frac{\vec{p}_{e1} \vec{p}_{e2}}{|\vec{p}_{e1}| |\vec{p}_{e2}|}$. Since the experiment is not able to distinguish between the first emitted and second emitted electron, the angle ranges from 0$^\circ$ to 180$^\circ$. 
The measured proton yield as a function of KER and $\alpha$ before applying any conditions on electron energy is shown in Fig.~\ref{fig1}(b).

\section{Electronic and nuclear dynamics in the KER-$\alpha$ representation\label{sec:simulations_KER_alpha}}

As discussed in the main manuscript, the measured fragmentation yield as a function of KER and $\alpha$ constitutes a powerful representation for depicting the coupled electron and nuclear motion taking place during the two ionization steps. 
In Fig.~2 of the main manuscript we compare the measured distribution to its  simulated classical expectation value for the case of vibrational motion taking place only on the \ssg{} energy level.
Here we provide details on these simulations and will furthermore show how the classical parametric curves look for different fragmentation pathways.

\begin{figure}[ht]
  \centering
  \includegraphics[width=0.5\columnwidth]{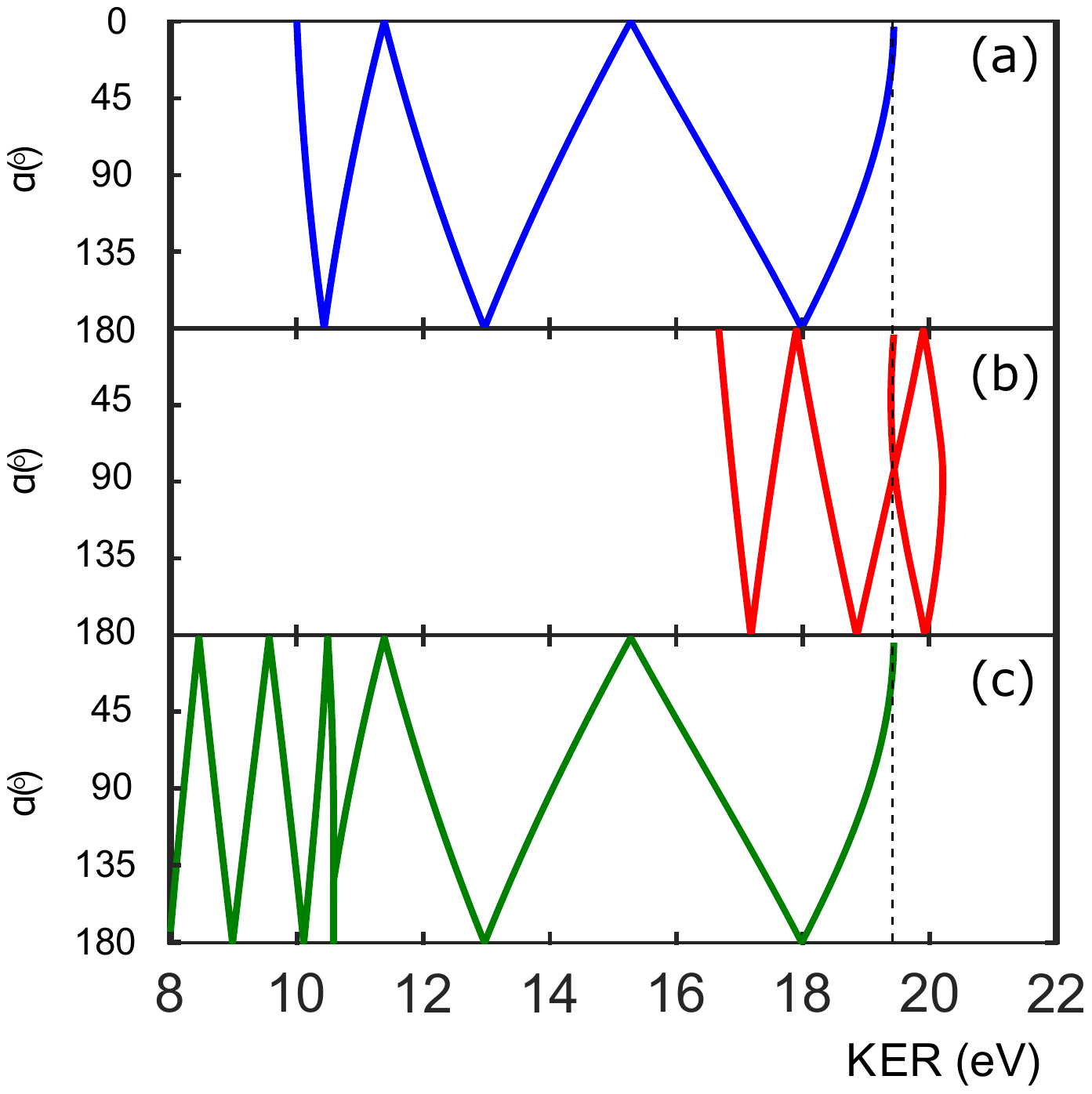}
  \caption{Classically calculated connection between the relative electron angle $\alpha$ and KER for selected fragmentation pathways. See text for details of the calculations. 
  (a) The nuclear wavepacket propagates on the \ssg{} state starting from the equilibrium internuclear distance of H$_2$. (b) Shakeup of the molecular ion to the \psu{} state immediately after the first ionization step. (c) Propagation on the \ssg{} state until the internuclear distance 2.6~a.u. where the \ssg{} and \psu{} states become resonant via a five photon transition and a transition to the \psu{} level is made. The dashed vertical line indicates the KER value corresponding to the equilibrium distance of H$_2$ (1.4~a.u.). }
  \label{fig2}
\end{figure}

Fig.~\ref{fig2} shows examples 
%of the classical expectation value of the fragmentation yield 
for these curves for three different fragmentation pathways. 
To calculate the parametric curves in Fig.~\ref{fig2} we solve Newton's equations on the \ssg{} and \psu{} potential energy curves given in Ref.~\citen{Sharp1970}. 
%Following the Franck-Condon principle, 
We assume that upon the first ionization step at the instant $t_1$ the \ssg{} curve of H$_2^+$ is populated at an internuclear distance $R$ given by the equilibrium distance of H$_2$. 
We further assume that the potential energy curve is unaffected by the laser electric field and that the nuclear motion is determined purely by the gradient of  the populated potential energy curve (\ssg{} or \psu{}) at any given internuclear distance.
Transitions due to absorption of photons between the \ssg{} and \psu{} curves after the first ionization step are modeled as vertical transitions. They can take place when the energies of the two levels $E_{2p\sigma_u}(R)$ and $E_{1s\sigma_g}(R)$ become resonant with $n$ photons, $E_{2p\sigma_u}(R)-E_{1s\sigma_g}(R)=n\hbar\omega$, or due to shake-up.  
The second ionization step takes place at an instant $t_2=t_1+\Delta t$. At $t_2$ the repulsive Coulomb curve is populated and the KER (in atomic units) is calculated as $\text{KER}=\frac{1}{R} + E_K$, with $E_K$ the kinetic energies of the protons at time $t_2$. 
Depending on the pathway of the fragmentation dynamics along the \ssg{} and \psu{} potential energy curves, the sequence of transitions between these two curves and the ionization delay $\Delta t$, the resulting KER-$\alpha$ parametric curves look distinctively different. Their shape, thus, is characteristic for the nuclear and electronic dynamics taking place during the time $\Delta t$. Comparison of the measured fragmentation yield resolved in the two-dimensional coordinate system KER vs. $\alpha$ to simulated curves, thus, allows identifying the molecular motion taking place during the interaction with the laser field.

The blue line in Fig.~\ref{fig2}(a) corresponds to the case, where vibrational motion taking place in between the two ionization steps at times $t_1$ and $t_2=t_1+\Delta t$, respectively, occurs exclusively on the \ssg{} energy level. 
The first ionization step starts the nuclear motion on the \ssg{} curve. If the second ionization step would occur immediately after the first one, $\Delta t \rightarrow 0$, the angle $\alpha$ between the momentum vectors of the two emitted electrons would be zero and one would measure a KER value indicated by the vertical dashed line in Fig.~\ref{fig2}(a) (larger than 19\,eV).
As $\Delta t$ increases, the nuclear motion proceeds on the \ssg{} level and the internuclear distance increases, resulting in decreasing KER. If the second ionization step would take place half a cycle after the first one, i.e., at $\Delta t=\pi/\omega$, with $\omega$ the laser oscillation frequency, the angle between the two electron momentum vectors would be $180^\circ$ and KER would amount to roughly 18\,eV. For $\pi/\omega < \Delta t < 2\pi/\omega$ the angle $\alpha$ decreases again towards zero and KER decreases further due to the ongoing nuclear stretch motion, until at $\Delta t = 2\pi/\omega$ the emission direction of the second electron equals that of the first one and the angle $\alpha$ becomes zero again. The nuclear stretch motion and the concomitant decrease of KER persist during the next two laser cycles. About three laser cycles after the first ionization step, at $\Delta t \approx 6\pi/\omega$ and  $\alpha\approx0$, the maximum classically allowed value of the internuclear distance $R$ for the assumed initial potential energy (dictated by the vertical transition to the \ssg{} curve at $R=1.4$\,a.u.) is reached. From there on, the internuclear distance decreases again and the blue curve in Fig.~\ref{fig2}(a) is traversed in the direction towards increasing KER.

The curve in Fig.~\ref{fig2}(b) corresponds to molecular dynamics where a shake-up process in the molecular ion to the \psu{} state immediately after the first ionization step takes place. From there on the vibrational wavepacket proceeds on the dissociative \psu{} energy curve. The molecule starts to dissociate and the internuclear distance $R$ quickly increases. As the \psu{} curve initially shows a steeper dependence on $R$ as the Coulomb curve, KER values higher than those obtained for sequential ionization with $\Delta t \rightarrow 0$ are obtained, as long as the second ionization step takes place within about the first laser cycle after the first one. Dissociation is completed within a delay $\Delta t$ corresponding to about three laser cycles. After this delay the internuclear distance has grown to such large values that the KER does not significantly decrease anymore for longer delays.

Fig.~\ref{fig2}(c), finally, shows a fragmentation scenario, where nuclear motion first proceeds on the \ssg{} energy curve, until at the internuclear distance 2.6~a.u., where the \ssg{} and \psu{} states become resonant via a five photon transition, a transition to the \psu{} level takes place (around KER$=10.3$\,eV). From this point on, the nuclear motion takes place on the \psu{} energy curve and the  molecule dissociates. As the gradient of the \psu{} energy curve is already quite small at and beyond the $R$-value where the transition happens, the nuclear stretch motion is slow and KER decreases only relatively little per laser cycle.

\section{Selection of ionization delay by gating on electron momentum \label{sec:simulations_ionization_dynamics_KER}}

We discuss in the main manuscript that gating on the electrons' energy can be used to select the range of relative delays between the two ionization steps, $\Delta t$, see Fig.~2 and the corresponding text in the main manuscript.
Here, we provide the results of simulations to visualize this capability.

\begin{figure}[t]
  \centering
  \includegraphics[width=0.75\columnwidth]{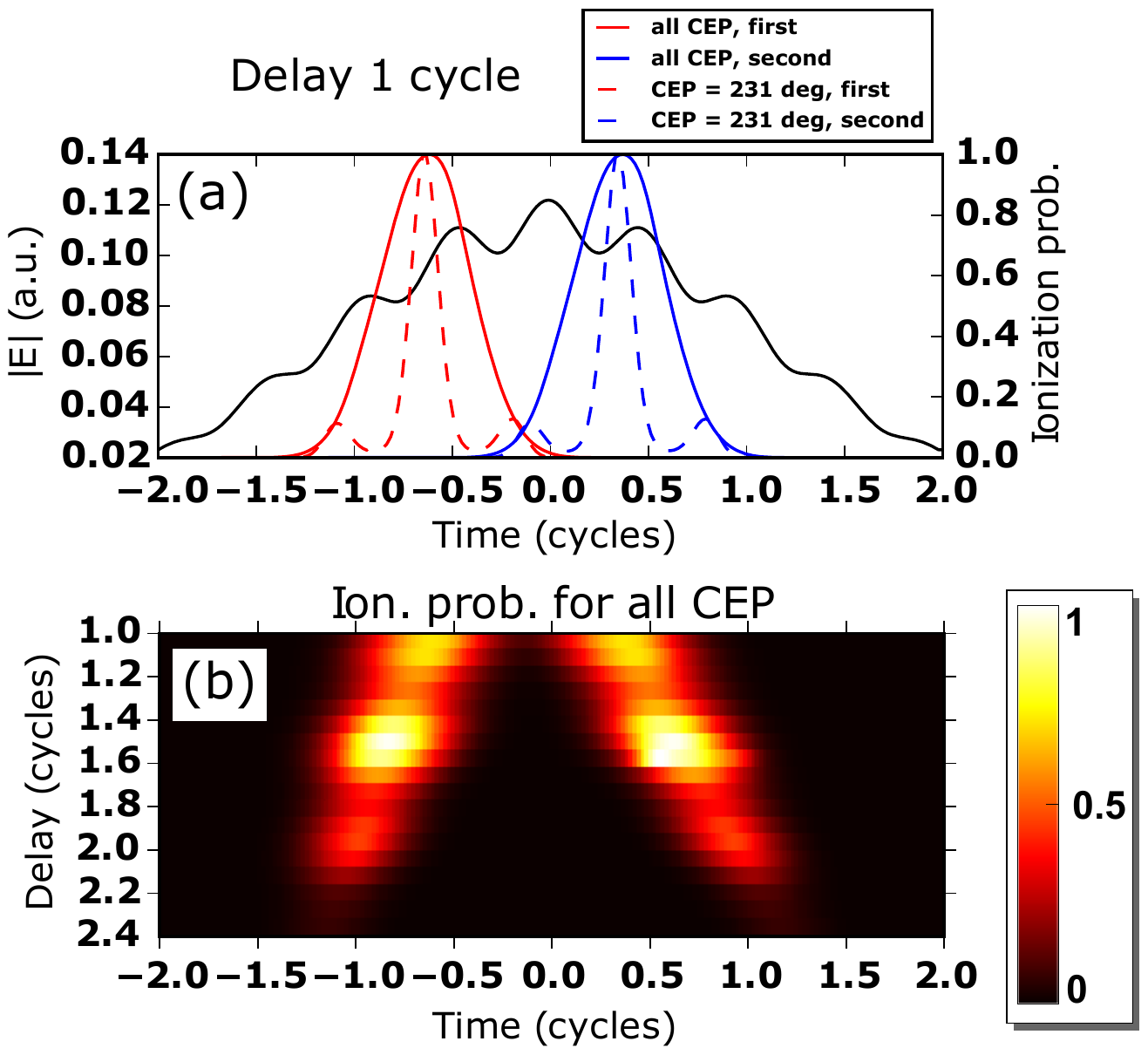}
  \caption{(a) Ionization rates of the first (dashed red) and second (dashed blue) ionization step  for the laser electric field shown in black. The corresponding full lines are obtained when one averages over all values of the CEP. (b) CEP-averaged ionization rates for the first (left branch) and second (right branch) emission event as a function of delay between the two events. All rates are normalized to one at their maximum values.}
  \label{fig3}
\end{figure}

\begin{figure}[t]
  \centering
  \includegraphics[width=0.55\columnwidth]{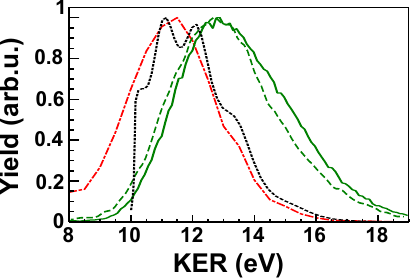}
  \caption{The full respectively dashed green lines show measured KER distributions of doubly ionized H$_2$ for peak intensities $9\times10^{14}$\,W/cm$^2$ and $1.2\times10^{15}$\,W/cm$^2$ for a pulse duration of $\approx 4.5$\,fs.
%   and an intensity of $9\times 10^{14}$\,W/cm$^2$. 
The  dashed-dotted red KER distribution is obtained for a 5.5-fs pulse with a peak intensity of $1.7\times10^{15}$\,W/cm$^2$. The dotted black KER distribution is obtained by simulations as described in Section~\ref{sec:simulations_ionization_dynamics_KER} for the  pulse parameters corresponding to the measured dashed-dotted red distribution.}
  \label{fig5}
\end{figure}

In these simulations we model the double ionization and nuclear dynamics using (semi-)classical models for a laser electric field of the form
\begin{equation}\label{equ:laserfield}
\vec{E}(t) = E_\parallel \cos(\omega t + \varphi) + E_\perp \sin(\omega t + \varphi),
\end{equation}
where $E_\parallel$ and $E_\perp = \varepsilon E_\parallel$ denote the peak electric field strengths along the major respectively minor polarization ellipse, $\varepsilon$ is the ellipticity, and $\varphi$ is the carrier-envelope phase (CEP).
For a comparison with the experiments, where the influence of the CEP is not considered, we repeat the simulations for different values of $\varphi \in [0,2\pi]$ and subsequently average over $\varphi$. The dependence of the laser electric field along the laser propagation direction (perpendicular to the $\parallel$ and $\perp$ directions) is neglected based on the thin jet used in the experiments (see Section~\ref{sec:exp_details}).

Ionization is modeled as described in Ref.~\citen{Tong2004} using an ADK-type formula. For the first ionization step, taking place at time $t_1$, the value of the ionization potential is $I_{p,1} = 15.4$ eV \cite{Dibeler1965}. The molecules are assumed to be randomly oriented in space and the angular dependence of the ionization potential is neglected. Ionization-depletion of the states is properly accounted for \cite{Tong2004}. 
The first ionization step triggers nuclear motion in the cation H$_2^+$.
This nuclear motion is modeled classically as described in Section~\ref{sec:simulations_KER_alpha}. 
The second ionization step happens $\Delta t$ after the first ionization step at time $t_2=t_1+\Delta t$, when the two nuclei have reached a certain internuclear distance, $R$. The ionization potential for the second ionization step, $I_{p,2}(R)$, depends  on $R$. 
$I_{p,2}(R)$ decreases monotonically with $R$ and could for example be determined by the energetic difference between the Coulomb explosion curve $1/R$ and the active molecular potential curve.
%for increasing internuclear distance of the cation H$_2^+$, $R$. 
We use the $I_{p,2}(R)$ curve as a parameter and adjust its offset and linear slope to obtain good agreement with the measured KER distribution. See the Fig.~\ref{fig1}(a) for a comparison of the calculated and measured KER distributions and the corresponding  $I_{p,2}(R)$ curve.
To obtain a KER distribution from our simulations we loop over the times $t_1$ and $t_2>t_1$ and calculate the double ionization probability $P=p_1 p_2$, where $p_1$ respectively $p_2$ denote the single and double ionization probabilities for this combination of $t_1$ and $t_2$ \cite{Tong2004}. The corresponding KER value is given by KER$=1/R$. By summing up all combinations of $t_1$ and $t_2>t_1$ with their respective double ionization probability $P$ we obtain distributions such as the one shown in Fig.~\ref{fig1}(a).

This model can be used to visualize the dependence between the two electrons' momenta $\vec{p}_{e1,e2}$ and the ionization delay $\Delta t$ that we exploit for obtaining Figs.~2(c) and 2(d) in the main manuscript.
Fig.~\ref{fig3} shows the ionization probability predicted by the model for different values of $\Delta t$, when the nuclear motion in between the two ionization steps takes place on the \ssg{} energy curve. 
Because the internuclear distance $R$ increases monotonically with $\Delta t$, the ionization potential for the second ionization step, $I_{p,2}(R)$, decreases with increasing $R$. To overcome the high ionization potential at small internuclear distances, the second ionization step at $t_2$ needs to happen as close as possible to the peak of the laser pulse where the electric field strength is highest. As a consequence, for small ionization delays $\Delta t$ (i.e., for small $R$), the ionization events will happen close to the pulse peak.
Indeed, Fig.~\ref{fig3} shows that for small delays the two emission events take place approximately symmetrically about the peak of the pulse, whereas for larger delays (larger $R$), the instants of ionization $t_1$ and $t_2$ shift symmetrically away from the pulse peak. 
The momenta of the two emitted electrons are given by $\vec{p}_{ei}=-\vec{A}(t_i), \quad i=1,2$ with  $\vec{A}(t)=-\int_{-\infty}^t \vec{E}(t') dt'$ derived from the laser electric field defined in Equ.~(\ref{equ:laserfield}). Thus, larger values of $\vec{p}_{e1,e2}$ are obtained for ionization times $t_1$ and $t_2$ close to the pulse peak. As a consequence, gating on the electrons' momenta, allows selecting the ionization delay $\Delta t$, cf. Fig.~\ref{fig3}.

%A separate calculation is made to obtain a relation between delay, inter-nuclear distance and KER. It is an integration that describes dynamics of nuclear stretch motion of H$_2^+$ cation and allows to estimate the distance between nuclei and its kinetic energy at certain delay. The nuclear wave-packet is represented as a point particle that moves on the energy potential. For every time delay quantities are evaluated like the angle between electrons -- $\alpha$, or final KER. This provides the $R(t_{\rm delay})$ mapping and finally also $\mathrm{KER}(t_{\rm delay})$ or $\mathrm{KER}(\alpha)$ which are used to calibrate measured figures. 

%{\old Similar effect that changes relative contributions in the KER distribution is the change of ionization potential in our model. If the ionization potential drops slower with nuclear stretching  higher KERs will be populated rather than lower ones. Obviously, the ionization potential is a property of the molecular ion and cannot be controlled.}

\section{Role of pulse duration in the molecular clock}

%We discuss the role of the laser pulse parameters in the molecular clock method.
%
We show in Fig.\,2 of the main manuscript and in the corresponding text that for the pulse parameters used in our experiment the useful range of $\Delta t$ is about $0.5$ to $3T$, with $T=2\pi/\omega$  the duration of the laser cycle. 
Smaller delays are suppressed due to the very large ionization potential for small internuclear distances, cf. the sketch in Fig.\,1(a) of the main manuscript. For obtaining larger delays (smaller KER values), our pulses are too short.
One could think that by simply increasing the laser intensity it would be possible to enhance the ionization yield at shorter internuclear distances, thereby accessing smaller values of $\Delta t$.
However, this is not the case since higher intensity would not only enhance  ionization around the pulse peak (where $\Delta t$ is small), but also at the rising and falling wings of the pulse envelope (corresponding to large values of $\Delta t$). 
Measurements for different peak intensity (but equal pulse duration) indeed show that the width of the KER distribution depends only marginally on intensity, compare the green curves in Fig.\,\ref{fig5}.

\begin{figure}[t]
  \centering
  \includegraphics[width=0.7\columnwidth]{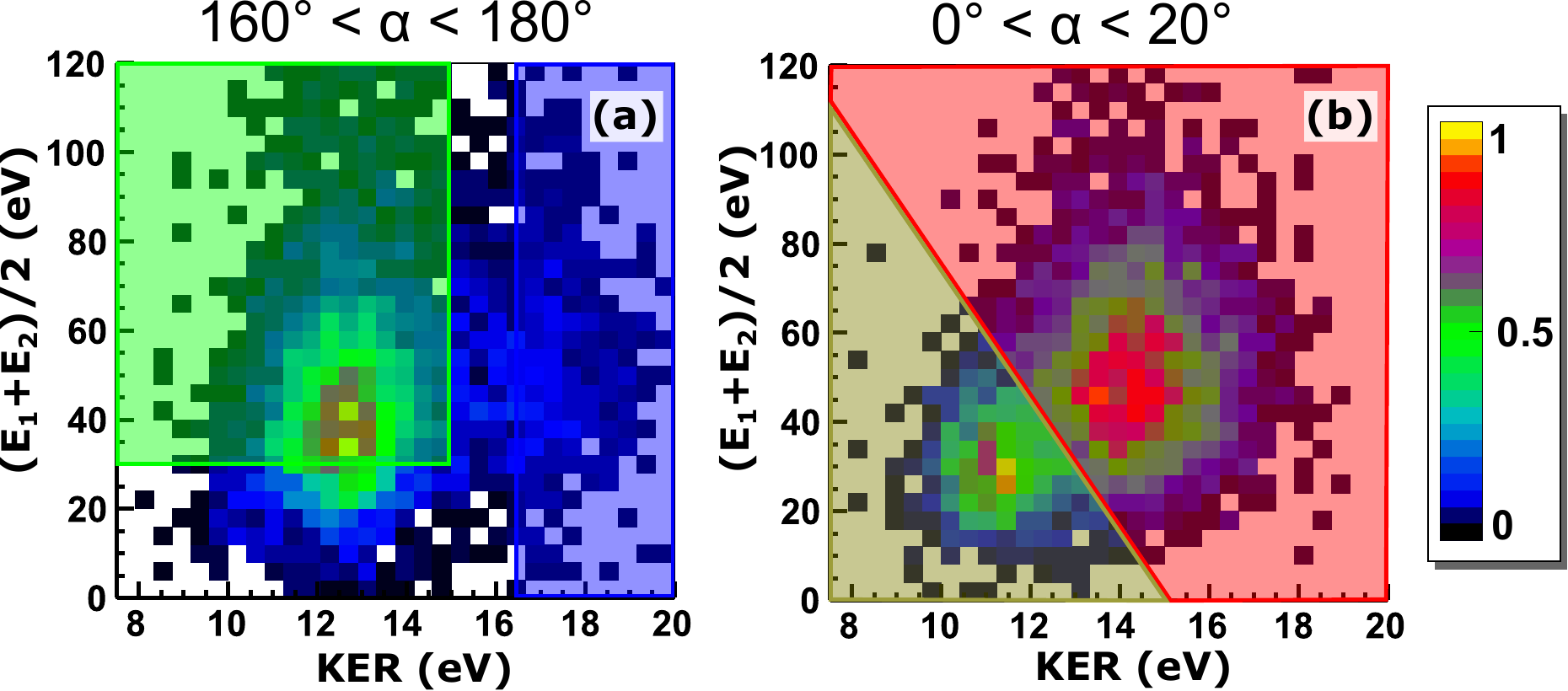}
  \caption{Measured distributions of the fragmentation yield as a function of KER and the averaged sum of the energy of the two emitted electrons for two ranges of the angle between the two electron momentum vectors $\alpha$ as indicated above the figures. The colored boxes visualize the selections from which the shapes of the vibrational wavepacket shown in Fig.~3 in the main manuscript is obtained by integration over the boxes. The colors blue, red, green and yellow correspond to the ionization delays $\Delta t=0.5T, 1.0T, 1.5T, 2.0T$, with $T=2\pi/\omega$ the laser cycle duration. %, marked \textbf{A, B, C} in Fig.~3 of the main manuscript.  respectively. 
}
  \label{fig4}
\end{figure}

In contrast, the KER distribution depends very sensitively on the duration of the pulse. An increase of only 1\,fs to 5.5\,fs results in a strong shift of the KER distribution to lower KER and therewith to larger values of $\Delta t$, as shown in Fig.\,\ref{fig5} by the measured (red) and simulated (black) KER distributions, where the simulated KER distribution is obtained by the procedure described in Section~\ref{sec:simulations_ionization_dynamics_KER}.
This high sensitivity is understandable, as the ionization potential for the second ionization step, $\textrm{H}_2^+ \rightarrow \textrm{H}_2^{2+}$, rapidly decreases with $R$, see the sketch in Fig.\,1(a) of the main manuscript. As a consequence, longer delays between the two ionization steps, i.e., larger $\Delta t$, are energetically preferred and will inevitably take place if they are accommodated by the pulse duration. Short delays (small $\Delta t$, high KER-values), in contrast, need to be enforced by squeezing the two ionization events into the short time-window of short pulses. Accessing $\Delta t$-values below $0.5T$ would necessitate intense pulses still shorter than 4.5\,fs used in the experiment described in the main manuscript.

\section{Retrieval of the vibrational wavepacket in H$_2^+$}

We show in Fig.~3 of the main manuscript that the vibrational wavepacket in H$_2^+$ launched during the first ionization step can be retrieved from measured data at certain instants during the pulse. As described in the text corresponding to that figure, the quality of the retrieval can be improved by careful choices of the ranges of KER, $\alpha$ and the absolute value of electron momentum respectively electron energy. Fig.\,\ref{fig4} visualizes the selection ranges that were used for the retrieval shown in Fig.~3 of the main manuscript.

%As presented in the main text of the article, KER distribution contains images of the nuclear wave-packet and it is possible to get its time evolution. We perform a retrieval of these images based on three quantities: relative angle $\alpha$, sum electron energy and electron energy difference. The relative angle $\alpha$ represents the time between the ionization steps. In the extremal values of $\alpha$, 0 or 180 degrees, we find contributions from times that correspond to even or odd multiples of laser half-cycle. But even in these extremal values of $\alpha$ the KER distribution of the neighbouring delays separated by full laser cycle are still overlapping. To separate them we use selections in sum electron energy as indicated in Fig. \ref{fig4}. This allows to unambiguously separate full cycle contributions. Thus, setting $\alpha$ to its extremal values, 0 and 180 degrees, allows to separate even from odd multiples of laser half-cycle.

%%%%%%%%%%%%%%%%%%%%%%%%%%%%%%%%%%%%%%%%%%%%%%%

\hspace{1cm}

%\bibliography{refs_markus}